\documentclass[aps,prd,preprint,groupedaddress]{revtex4}

\usepackage{epsfig}

\begin{document}

\def\lp{\left. }
\def\rp{\right. }
\def\lr{\left( }
\def\rr{\right) }
\def\le{\left[ }
\def\re{\right] }
\def\lg{\left\{ }
\def\rg{\right\} }
\def\lb{\left| }
\def\rb{\right| }

\def\go{\tilde{g}}
\def\mg{m_{\go}}

\def\tg{t_{\go}}
\def\ug{u_{\go}}

\def\ts{t_{\tilde{q}}}
\def\us{u_{\tilde{q}}}

\def\msQ {m_{\tilde{Q}}}
\def\msD {m_{\tilde{D}}}
\def\msU {m_{\tilde{U}}}
\def\msS {m_{\tilde{S}}}
\def\msC {m_{\tilde{C}}}
\def\msB {m_{\tilde{B}}}
\def\msT {m_{\tilde{T}}}
\def\msusy{m_{\rm SUSY}}

\def\sq  {\tilde{q}}
\def\sql {\tilde{q}_L}
\def\sqr {\tilde{q}_R}
\def\ms  {m_{\sq}}
\def\msql{m_{\tilde{q}_L}}
\def\msqr{m_{\tilde{q}_R}}

\def\st  {\tilde{t}}
\def\stl {\tilde{t}_L}
\def\str {\tilde{t}_R}
\def\mstl{m_{\stl}}
\def\mstr{m_{\str}}
\def\sta {\tilde{t}_1}
\def\stb {\tilde{t}_2}
\def\msta{m_{\sta}}
\def\mstb{m_{\stb}}
\def\thst{\theta_{\tilde{t}}}

\def\sb  {\tilde{b}}
\def\sbl {\tilde{b}_L}
\def\sbr {\tilde{b}_R}
\def\msbl{m_{\sbl}}
\def\msbr{m_{\sbr}}
\def\sba {\tilde{b}_1}
\def\sbb {\tilde{b}_2}
\def\msba{m_{\sba}}
\def\msbb{m_{\sbb}}
\def\thsb{\theta_{\tilde{b}}}

\newcommand{\sfa}{\tilde{f}}

\newcommand{\SLASH}[2]{\makebox[#2ex][l]{$#1$}/}
\newcommand{\kslash}{\SLASH{k}{.15}}
\newcommand{\pslash}{\SLASH{p}{.2}}
\newcommand{\qslash}{\SLASH{q}{.08}}

\def\d  {{\rm d}}
\def\eps{\varepsilon}
\def\O  {{\cal O}}

\def\beq{\begin{equation}}
\def\eeq{\end{equation}}
\def\bea{\begin{eqnarray}}
\def\eea{\end{eqnarray}}

\preprint{hep-ph/0303058}
\title{Gluino Pair Production in \boldmath$e^+e^-$ and \boldmath$\gamma\gamma$
 Collisions at CERN CLIC}
\author{Stefan Berge}
\affiliation{{II.} Institut f\"ur Theoretische Physik, Universit\"at Hamburg,
             Luruper Chaussee 149, D-22761 Hamburg, Germany}
\author{Michael Klasen}
\email[]{michael.klasen@desy.de}
\affiliation{{II.} Institut f\"ur Theoretische Physik, Universit\"at Hamburg,
             Luruper Chaussee 149, D-22761 Hamburg, Germany}
\date{\today}
\begin{abstract}
We confront the generally small cross sections for gluino pair production in
$e^+e^-$ annihilation with the much larger ones in photon-photon scattering at
a multi-TeV linear collider like CERN CLIC. The larger rates and the steeper
rise of the cross section at threshold may allow for a precise gluino mass
determination in high-energy photon-photon collisions for a wide range of
squark masses and post-LEP SUSY benchmark points.
\vfill
\centerline{\bf To appear in the CERN Yellow Report from the CLIC Physics Study Group.}
\vfill
\end{abstract}
\maketitle

Weak-scale supersymmetry (SUSY) is one of the most attractive extensions of the
Standard Model (SM) of particle physics. If it is realized in nature, SUSY
particles will be discovered either at Run II of the Fermilab Tevatron or
within the first years of running at the CERN LHC. Reconstruction of the SUSY
Lagrangian and a precise determination of its free parameters will, however,
require the clean environment of a linear $e^+e^-$ collider, where in
particular the masses, phases, and (electroweak) couplings of sfermions and
gauginos will be determined with high accuracy. However, the mass and coupling
of the gluino will pose some difficulties, since the gluino couples only to
strongly interacting particles and is thus produced only at the one-loop level
or in multi-parton final states. In a recent publication, we have investigated
gluino pair production through triangular quark/squark loops in $e^+e^-$
annihilation with energies up to 1 TeV, which may become available in the
nearer future at linear colliders like DESY TESLA \cite{Berge:2002ev}.
Due to large cancellation effects, we found that promisingly large cross
sections can only be expected for scenarios with large left-/right-handed
up-type squark mass splittings or with large top-squark mixing and for gluino
masses up to 500 GeV. Small gluino masses of 200 GeV might be measured with a
precision of about 5 GeV in center-of-mass energy scans with luminosities of
100 fb$^{-1}$/point.

In this Report, we focus on a multi-TeV linear collider like CERN CLIC with
electron (positron) beam polarization of 80\% (60\%) and integrated luminosity
of 1000 fb$^{-1}$/year and compare the generally small production rates in
$e^+e^-$ annihilation to the larger ones in photon-photon scattering. Further
details on gluino pair production in high-energy photon collisions can be
found in Ref.\ \cite{Berge:2003cj}. In the photon collider version of CERN
CLIC, 100\% polarized laser photons are backscattered from two electron beams,
whose helicities must be opposite to those of the laser photons in order to
maximize the number of high-energy photons. In the strong fields of the laser
waves, the electrons (or high-energy photons) can interact simultaneously with
several laser photons. This non-linear effect increases the threshold parameter
for $e^+e^-$ pair production to $X=(2+\sqrt{8})(1+\xi^2)\simeq6.5$, which
corresponds to laser wavelengths of 4.4 $\mu$m at an $e^+e^-$ center-of-mass
energy of 3 TeV \cite{Burkhardt:2002vh}. The maximal fractional energy of the
high-energy photons is then $x_{\max}=X/(X+1)=0.8\bar{6}$. Since the low-energy
tail of the photon spectrum is neither useful nor well understood, we use only
the high-energy peak with $x>0.8\,x_{\max}=0.69\bar{3}$ and normalize our cross
sections such that the expected number of events can be obtained through simple
multiplication with the envisaged photon-photon luminosity of 100-200
fb$^{-1}$/year. This requires reconstruction of the total final-state energy,
which may be difficult due to the missing energy carried away by the
escaping lightest SUSY particles (LSPs). However, high-energy
photon collisions allow for cuts on the relative longitudinal energy in
addition to the missing-$E_T$ plus multi-jet, top or bottom (s)quark, and/or
like-sign lepton analyses performed at hadron colliders, and sufficiently
long-lived gluinos can be identified by their typical $R$-hadron signatures.

We adopt the current mass limit $\mg\geq 200$ GeV from the CDF
\cite{Affolder:2001tc} and D0 \cite{Abachi:1995ng} searches in the jets with
missing energy channel, relevant for non-mixing squark masses of $\ms\geq 325$
GeV and $\tan\beta = 3$. Since values for the ratio of the Higgs vacuum
expectation values, $\tan\beta$, below 2.4 are already excluded by the CERN LEP
experiments, and since values between 2.4 and 8.5 are only allowed in a very
narrow window of light Higgs boson masses between 113 and 127 GeV
\cite{lhwg:2001xx}, we employ a safely high value of $\tan\beta = 10$.
For a conservative comparison of the $\gamma\gamma$ and $e^+e^-$ options at
CERN CLIC, we maximize the $e^+e^-$ cross section by adopting the smallest
allowed universal squark mass of $\ms\simeq\msusy=325$ GeV and large top-squark
mixing with $\thst=45.195^\circ$, $\msta=110.519$ GeV, and $\mstb=505.689$ GeV,
which can be generated by choosing appropriate values for the Higgs mass
parameter, $\mu=-500$ GeV, and the trilinear top-squark coupling, $A_t=648.512$
GeV \cite{Hahn:2001rv}. The SUSY one-loop contributions to the $\rho$-parameter
and the light top-squark mass $\msta$ are then still significantly below and
above the CERN LEP limits, $\rho_{\rm SUSY} < 0.0012^{+0.0023}_{-0.0014}$ and
$\msta\geq 100$ GeV \cite{Hagiwara:pw,lswg:2002xx}. For small values of
$\tan\beta$, mixing in the bottom squark sector remains small, and we take
$\thsb=0^\circ$. A full set of Feynman diagrams can be generated and evaluated
with the computer algebra packages FeynArts \cite{Hahn:2000kx} and FormCalc
\cite{Hahn:1998yk}.

\begin{figure}
 \centering
 \includegraphics*[width=0.49\columnwidth]{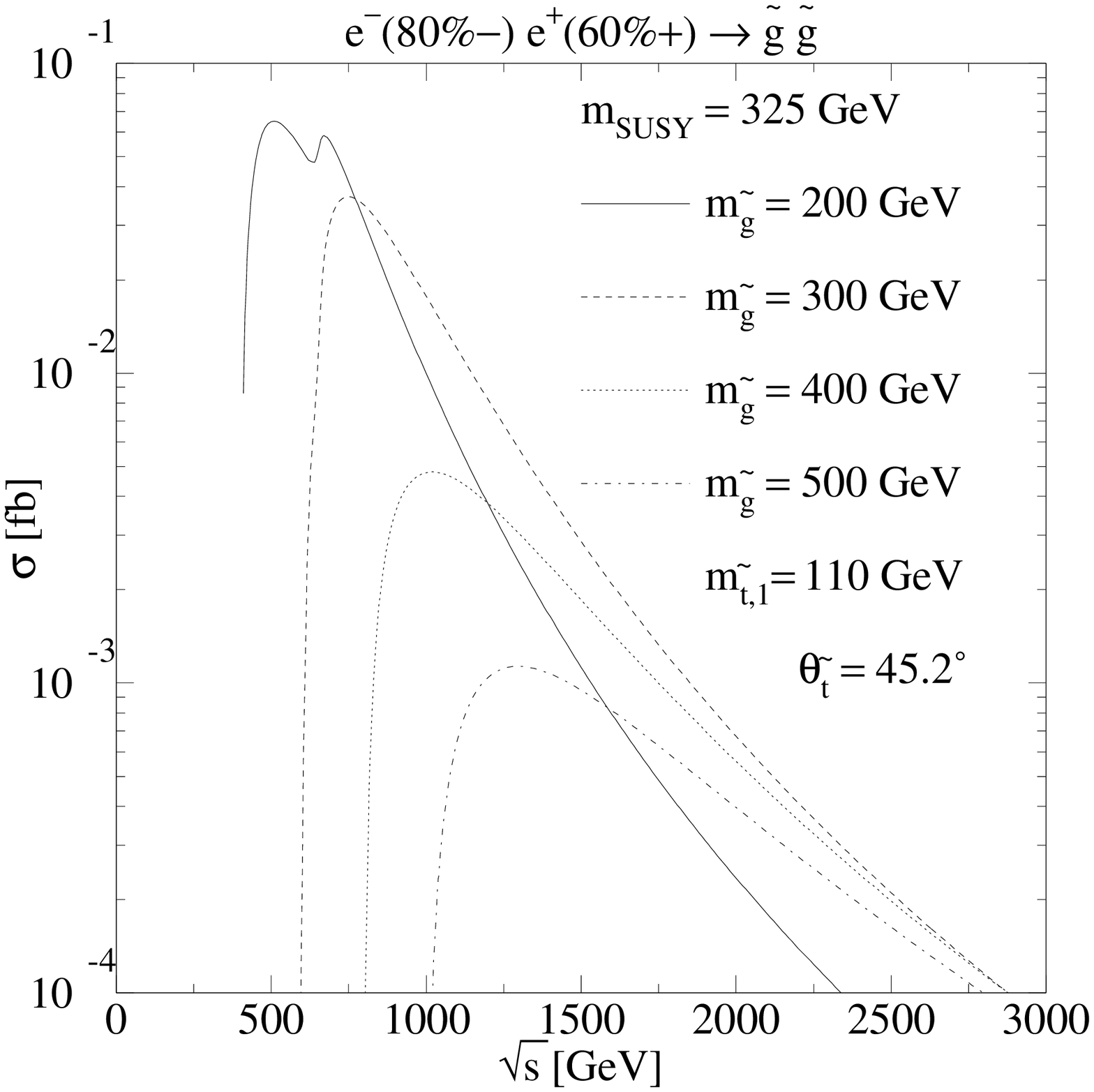}
 \includegraphics*[width=0.49\columnwidth]{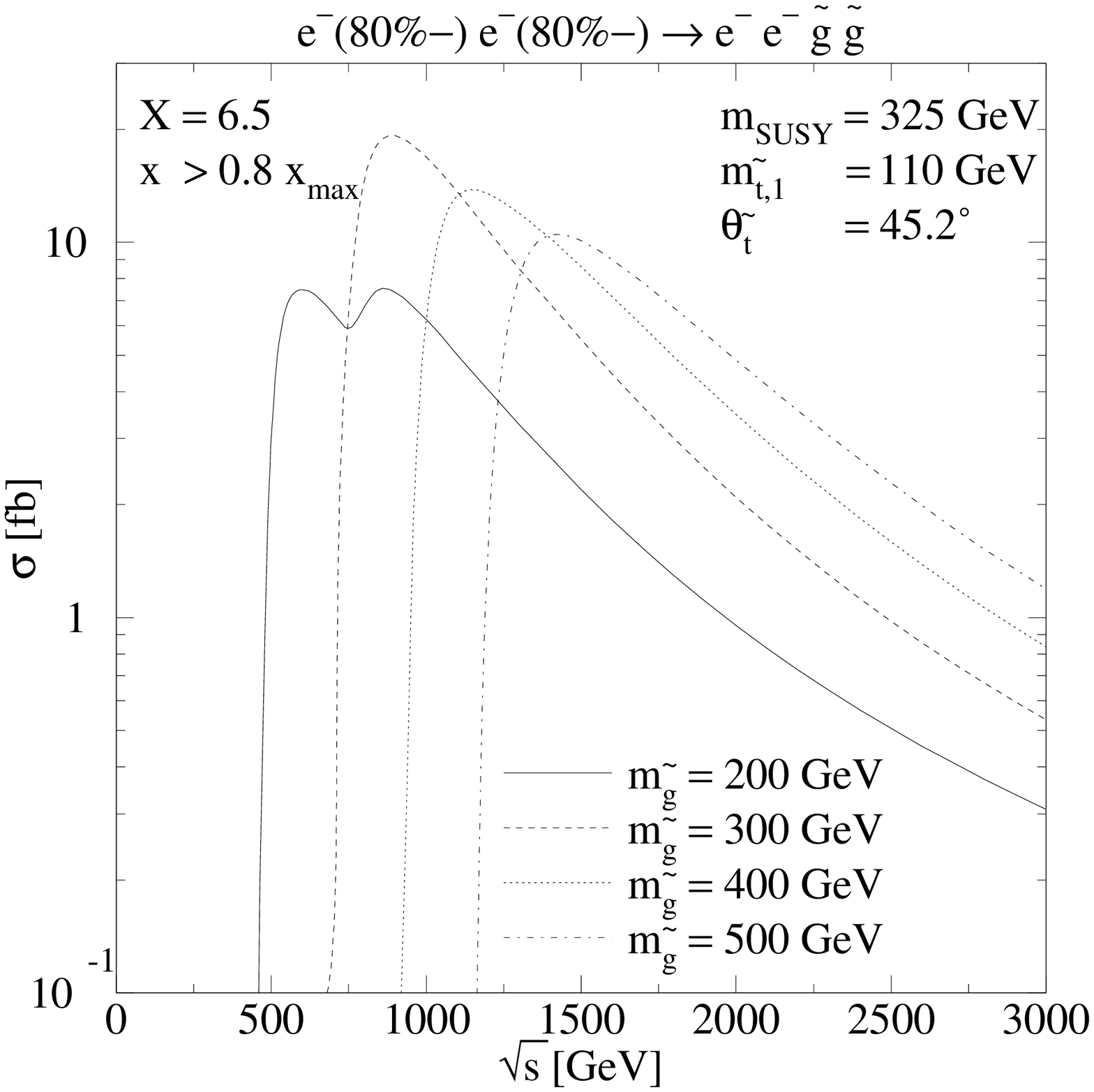}
 \caption{\label{fig:1}Gluino pair production cross sections in $e^+e^-$
 annihilation (left) and $\gamma\gamma$ collisions (right) as a function of the
 $e^\pm e^-$ center-of-mass energy and for various gluino masses. The
 photon-photon luminosity has been normalized to unity in the high-energy
 peak.}
\end{figure}

Fig.\ \ref{fig:1} shows a comparison of the total cross sections expected in
%
%
$e^+e^-$ annihilation and $\gamma\gamma$ collisions for gluino masses between
200 and 500 GeV. A light squark mass $\msusy$ and large top-squark mixing has
been chosen in order to maximize the $e^+e^-$ cross section. Neveretheless, it
stays below 0.1 fb and falls steeply with $\mg$, so that gluino pair production
will be unobservable for $\mg>500$ GeV irrespective of the collider energy. In
contrast, the $\gamma\gamma$ cross section reaches around ten fb for a wide
range of $\mg$. In $e^+e^-$ annihilation the gluinos are produced as a
$P$-wave and the cross section rises rather slowly, whereas in $\gamma\gamma$
collisions they can be produced as an $S$-wave and the cross section rises
much faster.

\begin{figure}
 \centering
 \includegraphics*[width=0.49\columnwidth]{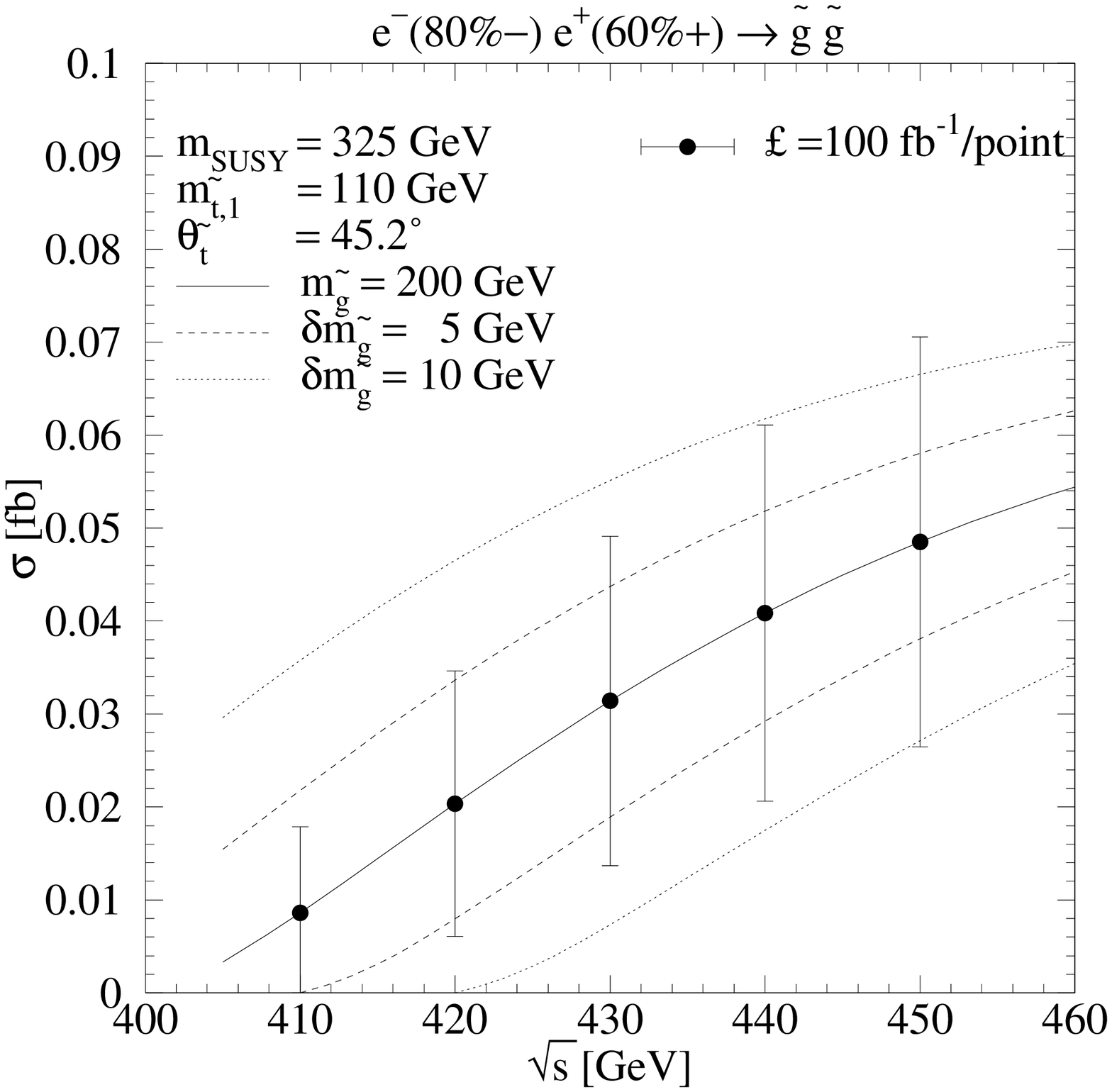}
 \includegraphics*[width=0.49\columnwidth]{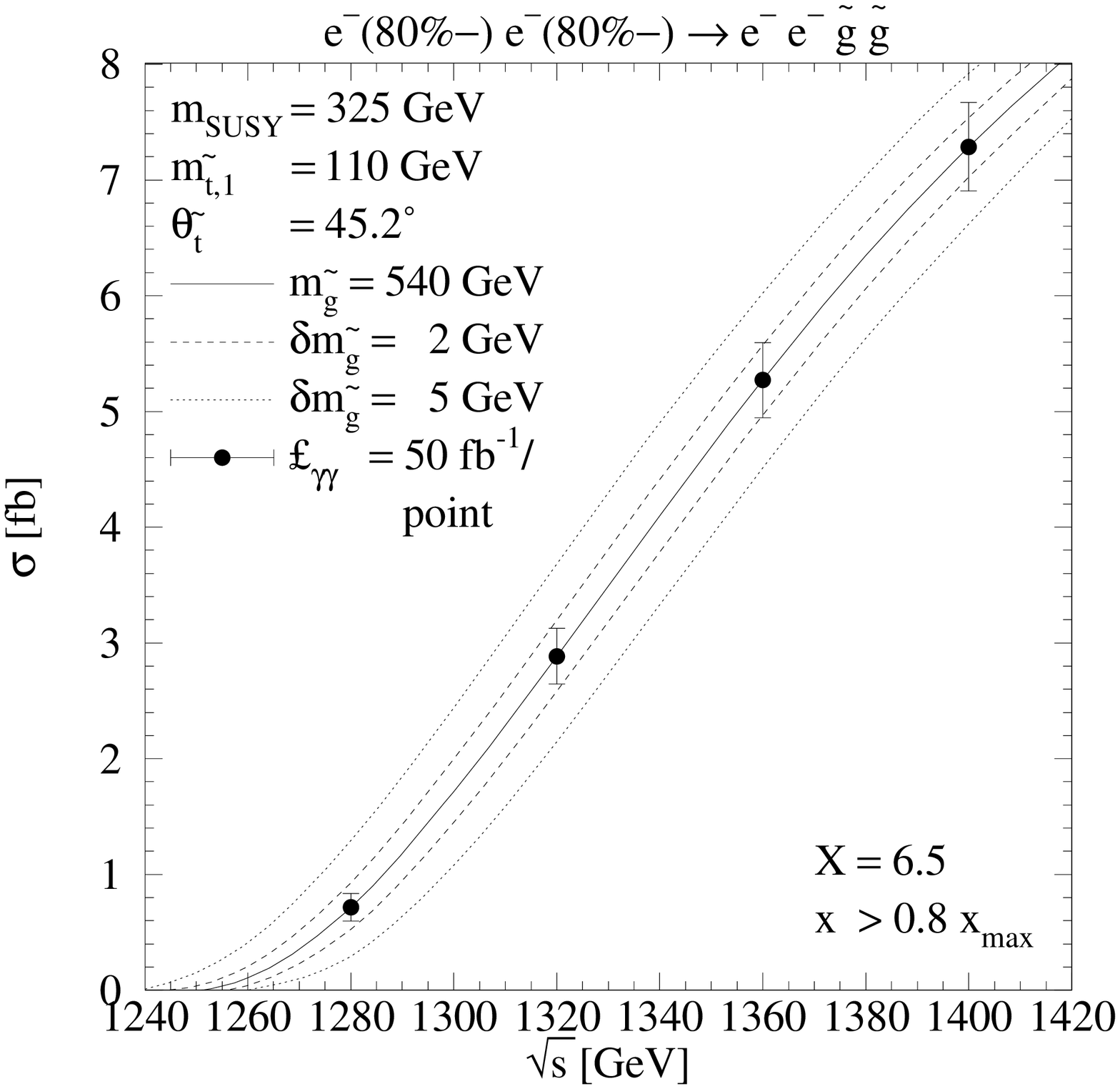}
 \caption{\label{fig:2}Sensitivity of the $e^+e^-$ annihilation (left) and
 $\gamma\gamma$ scattering cross section (right) to the mass of the
 pair-produced gluino. The photon-photon luminosity has been normalized to
 unity in the high-energy peak.}
\end{figure}

The different threshold behavior can be observed even more clearly in Fig.\
\ref{fig:2}, where the sensitivities of $e^+e^-$ and $\gamma\gamma$ colliders
%
%
to the gluino mass are compared. For the CERN LHC experiments, a precision of
$\pm30\,...\,60$ ($12\,...\,25$) GeV is expected for gluino masses of 540
(1004) GeV \cite{:1999fr,Abdullin:1998pm}. If the masses and mixing angle(s)
of the top (and bottom) squarks are known, a statistical precision of
$\pm5\,...\,10$ GeV can be achieved in $e^+e^-$ annihilation for $\mg=200$ GeV
for an integrated luminosity of 100 fb$^{-1}$ per center-of-mass energy point.
A precision of $\pm2\,...\,5$ GeV may be obtained at a CERN CLIC photon
collider for $\mg=540$ GeV and an integrated photon-photon luminosity of 50
fb$^{-1}$ per point, provided that the total final-state energy can be
sufficiently well reconstructed. Of course, uncertainties from a realistic
photon spectrum and the detector simulation add to the statistical error.

\begin{figure}
 \centering
 \includegraphics*[width=0.9\columnwidth]{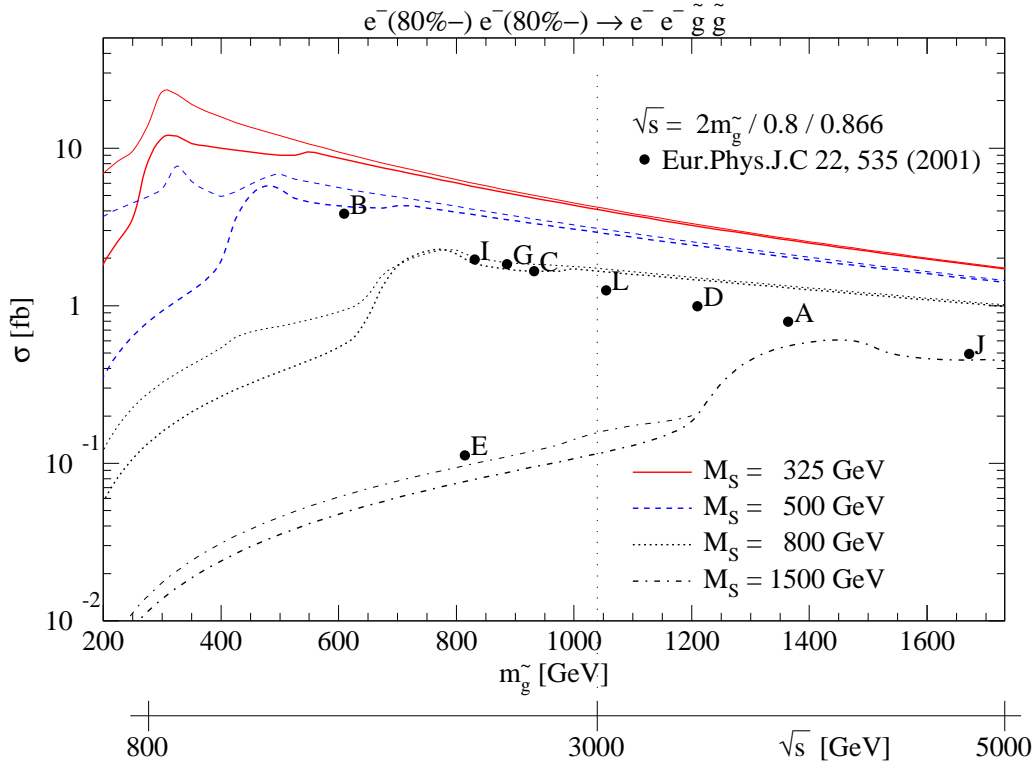}
 \caption{\label{fig:3}Dependence of the gluino pair production cross section
 in $\gamma\gamma$ collisions on the universal squark mass $\msusy$ for
 no squark mixing (thick curves) and maximal top-squark mixing (thin curves).
 The photon-photon luminosity has been normalized to unity in the high-energy
 peak. Also shown are the cross sections for the SUSY benchmark points of Ref.\
 \cite{Battaglia:2001zp}.}
\end{figure}

Finally, we demonstrate in Fig.\ \ref{fig:3} that gluino pair production
%
%
in $\gamma\gamma$ collisions depends only weakly on the universal squark mass
$\msusy$ and even less on the top-squark mixing. This is in sharp contrast to
the results obtained in $e^+e^-$ annihilation \cite{Berge:2002ev}. In this
plot, the $e^-e^-$ center-of-mass energy is chosen close to the threshold for
gluino pair production and is varied simultaneously with $\mg$. Also shown in
Fig.\ \ref{fig:3} are several post-LEP SUSY benchmark points, which have
recently been proposed within the framework of the constrained MSSM
\cite{Battaglia:2001zp}. Studies similar to those performed in Fig.\
\ref{fig:2} show that with the exception of point E, where only about ten
events per year are to be expected, the gluino mass can be dermined with a
precision of $\pm20$ GeV (point J) or better.

In conclusion, the reconstruction of the SUSY Lagrangian and the precise
determination of its free parameters are among the paramount objectives of any
future linear $e^+e^-$ collider. Determination of the gluino mass and coupling
will, however, be difficult, since the gluino couples only strongly and its
pair production cross section suffers from large cancellations in the
triangular quark/squark loop diagrams. A photon collider may therefore be
the only way to obtain precise gluino mass determinations and visible gluino
pair production cross sections for general squark masses and would thus
strongly complement the physics program feasible in $e^+e^-$ annihilation.



\end{document}